%% file: main.tex
\documentclass{article}
\PassOptionsToPackage{numbers, compress}{natbib}
\usepackage[final,preprint]{neurips_2024}

\usepackage[utf8]{inputenc} 
\usepackage[T1]{fontenc}    
\usepackage{hyperref}       
\usepackage{url}            
\usepackage{booktabs}       
\usepackage{amsfonts}       
\usepackage{nicefrac}       
\usepackage{microtype}      
\usepackage[dvipsnames]{xcolor}         
\usepackage{amsmath}
\usepackage{cleveref}
\usepackage{graphicx}
\usepackage{caption}
\usepackage{subcaption}

\input{math_commands.tex}

\hypersetup{
    colorlinks,
    colorlinks = true,
    linkcolor = OliveGreen,
    urlcolor  = Cerulean,
    citecolor = MidnightBlue,
    anchorcolor = blue
}

\title{Deconstructing equivariant representations in molecular systems}
\author{Kin Long Kelvin Lee \quad Mikhail Galkin \quad Santiago Miret \\ Intel Labs}
\date{August 2024}

\begin{document}

\maketitle

\begin{abstract}
Recent equivariant models have shown significant progress in not just chemical property prediction, but as surrogates for dynamical simulations of molecules and materials. Many of the top performing models in this category are built within the framework of tensor products, which preserves equivariance by restricting interactions and transformations to those that are allowed by symmetry selection rules. Despite being a core part of the modeling process, there has not yet been much attention into understanding what information persists in these equivariant representations, and their general behavior outside of benchmark metrics. In this work, we report on a set of experiments using a simple equivariant graph convolution model on the QM9 dataset, focusing on correlating quantitative performance with the resulting molecular graph embeddings. Our key finding is that, for a scalar prediction task, many of the irreducible representations are simply ignored during training---specifically those pertaining to vector ($l=1$) and tensor quantities ($l=2$)---an issue that does not necessarily make itself evident in the test metric. We empirically show that removing some unused orders of spherical harmonics improves model performance, correlating with improved latent space structure. We provide a number of recommendations for future experiments to try and improve efficiency and utilization of equivariant features based on these observations.
\end{abstract}

\section{Introduction}

Equivariant modeling has been known and subsequently shown to be extremely useful for holistic descriptions of atomic systems by respecting native symmetries of both systems (i.e. molecules) and the operators that act on them for property modeling. The core concept of equivariance lends itself to enabling a new degree of research productivity in applications ranging from molecular conformer generation \citep{satorras2021n, thiede2022conformer} to condensed phase materials discovery \citep{merchant2023scaling, miret2023the, miret2024perspective}. One of the most powerful applications of equivariant models is in the dynamical simulation of materials via the creation of interatomic potentials, which contrasts conventional atomic force-fields by allowing a single neural network architecture to support a wide range of elements of the periodic table, as opposed to bespoke force fields tuned for particular material classes. Recent examples of equivariant neural networks developed for this purpose are the SE(3)-Transformer~\citep{fuchs2020se}, SEGNN~\citep{brandstetter2021geometric}, NequIP~\citep{batzner2022nequip}, MACE~\citep{batatia2022mace}, eSCN~\citep{escn}, and Equiformer~\citep{liao2022equiformer,equiformer_v2} amongst others \citep{duval2023hitchhiker}. While many of these architectures have been successful in property modeling for a diversity of chemical systems, further work remains in understanding their capabilities and limitations as machine learning interatomic potentials \citep{bihani2024egraffbench}.

The common theme between these architectures is how they achieve equivariance; tensor products using spherical harmonic functions as a basis provides a unifying framework covering invariance to specific types of equivariance, primarily through the order $l$ and parity of spherical harmonics used. This framework was originally devised for analyzing how different sources of angular momenta can couple together \citep{condonTheoryAtomicSpectra1991,zareAngularMomentumUnderstanding1988}, and equivariant neural networks that use spherical harmonics and tensor products follow the same guiding principles: symmetry rules dictate which sets of irreducible representations can interact, and their coupling strength is given by the Clebsch-Gordon coefficients. Despite their general success, current state-of-the-art models like MACE \citep{batatia2022mace} limit their order of spherical harmonics to $l{\sim}2$, and in the case of NequIP, $l=3$ \citep{batzner2022nequip}. From a signal processing perspective, higher order spherical harmonics can be thought of as analogous to high frequency components in Fourier series; their complex node structure is potentially capable of encoding correspondingly complex atomic motifs, in addition to providing the same equivariant preserving treatment to the prediction of higher rank properties as the need arises, such as octopole (a rank-three tensor) moments and multiphoton spectroscopy (rank-$k$ for $k\leq$ the number of photons) \cite{WESTERN2017221}.

Recent work by \citet{leeScalingComputationalPerformance2024} investigated approaches to improve the computational efficiency of spherical harmonics by implementing them in the Triton language \citep{tillet2019triton}. The aim of \citet{leeScalingComputationalPerformance2024} was to unblock research avenues into using higher orders of spherical harmonics---and therefore richer feature spaces---by relieving their computational complexities. In this paper, we built on top of \citet{leeScalingComputationalPerformance2024} to provide a new set of implementations for spherical harmonics up to $l=10$, described in \Cref{sec:equitriton}, which remove inter-order dependencies in spherical harmonics. This reformulation allows free, efficient composition of spherical harmonics. We apply these kernels in a series of experiments that seek to understand how irreducible representations ultimately affect regression tasks. Our experiments in \Cref{sec:results} show that equviariance-preserving feature sets can potentially be left unused, leading to worse model performance. The experimental space allowed us to correlate quantiative metrics with qualitative visualizations of the latent embedding space, which can be correlated to poorer test errors. Finally, we propose testable hypotheses and recommendations for future investigations into the optimal usage and design of tensor product models.

\section{Methodology}

In this work, we implement a highly simplified equivariant graph convolution model inspired by NequIP \citep{batzner2022nequip} to embed molecular graphs. In an attempt to arrive at semantic embeddings, we train this architecture on atomization energy prediction at 0\,K ($U^\mathrm{atom}_0$) using the QM9 dataset \citep{ramakrishnanQuantumChemistryStructures2014a,ruddigkeitEnumeration166Billion2012}: at a high level, the model couples (initially) scalar atom features of dimension $h$ with atom positions embedded in a \emph{basis set} of $L$\footnote{We use $L$ to denote a set of spherical harmonics, and $l$ to refer to individual orders and subsets.} spherical harmonics with fully connected tensor products to yield node embeddings with a dimensionality of $d = \sum^L_{l \in L} h (2l + 1)$, where $i$ indexes the orders of spherical harmonics specified, and $l \in \{0-10\}$ in our current implementation. A sequence of graph convolutions aggregate information across multiple node hops \emph{and} irreducible representations, with the latter according to allowed tensor product coupling schemes. After convolution, we use a scalar ($l=0$) projection to obtain node-wise contributions of the energy, with the total atomization energy given as the sum over nodes.

Next, we use PHATE \citep{moonVisualizingStructureTransitions2019} as a method of obtaining low-dimensional projections of the embeddings. PHATE preserves local and global structure of the high dimensional manifold by first converting euclidean distances into a affinity/likelihood through a modified Gaussian kernel transformation, and subsequently represents pairwise distances in the manifold as marginalized likelihoods---intuitively, ``minimum energy paths'' between points---which are projected down to lower dimensions. Consequently, PHATE is able to faithfully represent both local \emph{and} global structure in the data, contrasting other similar approaches like $t$-SNE \citep{JMLR:v9:vandermaaten08a} and traditional principal components analysis. We apply PHATE to the joint ($d$ dimensionality) graph embeddings as well as features that correspond with each irreducible representation---this decomposition permits insight into how equivariant graph networks use individual sets of [$h(2l+1)$] tensor features for this particular scalar prediction task. To guide interpretation of these projections, points are colored based on a straightforward measure of chemical complexity; we use the normalized spacial score (NSPS) \citep{krzyzanowskiSpacialScoreComprehensive2023} (see \Cref{sec:nsps} for details). Training/inference code and implementations of the Triton spherical harmonics can be found at the \href{https://github.com/IntelLabs/EquiTriton/commit/728b10719758bb012710196f677d9f3ce7aefefa}{EquiTriton repository}.

\section{Results \& Discussion} \label{sec:results}

\subsection{Quantitative metrics}

Table \ref{tab:modelperformance} presents the test-set performance for a number of experiment configurations, split into two parts. In the top half, the best performing hyperparameters correspond to the ``canonical'' equivariant set, $L=[0,1,2]$, which is also the smallest configuration in terms of the number of parameters. The remaining runs in this part of the table add higher order spherical harmonics on an individual basis, and does not indicate any strong correlation with test error, performing slightly worse than the canonical set. 

In the second half of the table, we constrained experimental parameters by decreasing the number of training epochs and the nominal size of the hidden dimension; there is a large dynamic range in time to completion due to the additional tensor contractions \emph{and} model size, which grows quickly with $L$. The fewer training epochs generally accounts for the poorer performance relative to experiments in the top half of Table \ref{tab:modelperformance}. In this set of experiments, the test performance interestingly improves, then degrades with higher order $l$---we initially postulate that this could be due to the substantially larger feature space\footnote{Since the number of projections scales as $2l+1$, parameters belonging to higher orders make up an increasing fraction of the full model.} leading to overparametrization and thus poorer convergence dynamics. Perhaps the most interesting result is in the last line, where we \emph{omit} $l=1,2$, which yields the best performing model of the set; despite having roughly the same number of learnable parameters as $L=[0,1,2,3,4]$, performs over two times better, and when trained for 100 epochs, outperform even the canonical $L=[0,1,2]$ by an order of magnitude. In the next section, we will attempt to rationalize these findings using the PHATE embedding projections.

\begin{table}[h]
    \begin{center}
    \caption{Test set performance for a small set of hyperparameters. Test error corresponds to the mean squared value in atomization energy in units of eV. Figures bolded denote the lowest test error within the subset of experiments.}
    \begin{tabular}{l l l c c }
        \toprule
         $L$ & Epochs & $h$ & \# of parameters (M) & Test error (eV) $\downarrow$ \\
         \midrule
         $[0,1,2]$ & 100 & 32 & 1.6 & \textbf{0.12} \\
         $[0,1,2,4]$ & 100 & 32 & 3.0 & 0.15 \\
         $[0,1,2,6]$ & 100 & 32 & 2.6 & 0.21 \\
         $[0,1,2,8]$ & 100 & 32 & 2.6 & 0.19 \\
         $[0,1,2,10]$ & 100 & 32 & 2.6 & 0.19 \\
         \midrule
         $[0,1,2,3,4]$ & 30 & 16 & 0.8 & 1.24 \\
         $[0,1,2,3,4]$ & 100 & 16 & 0.8 & 0.22 \\
         $[0,1,2,3,4,5,6]$ & 30 & 16 & 1.9 & 0.73 \\
         $[0,1,2,3,4,5,6,7,8]$ & 30 & 16 & 3.7 & 1.02  \\
         $[0,3,4,5,6]$ & 30 & 16 & 0.8 & 0.52 \\
         $[0,3,4,5,6]$ & 100 & 16 & 0.8 & \textbf{0.01} \\
         \bottomrule
    \end{tabular}
    \label{tab:modelperformance}
    \end{center}
\end{table}


\subsection{Embedding projections}
\label{sec:embedding-projection}

\begin{figure}
    \centering
    \begin{subfigure}[c]{0.9\textwidth}
        \centering
        \includegraphics[width=\linewidth]{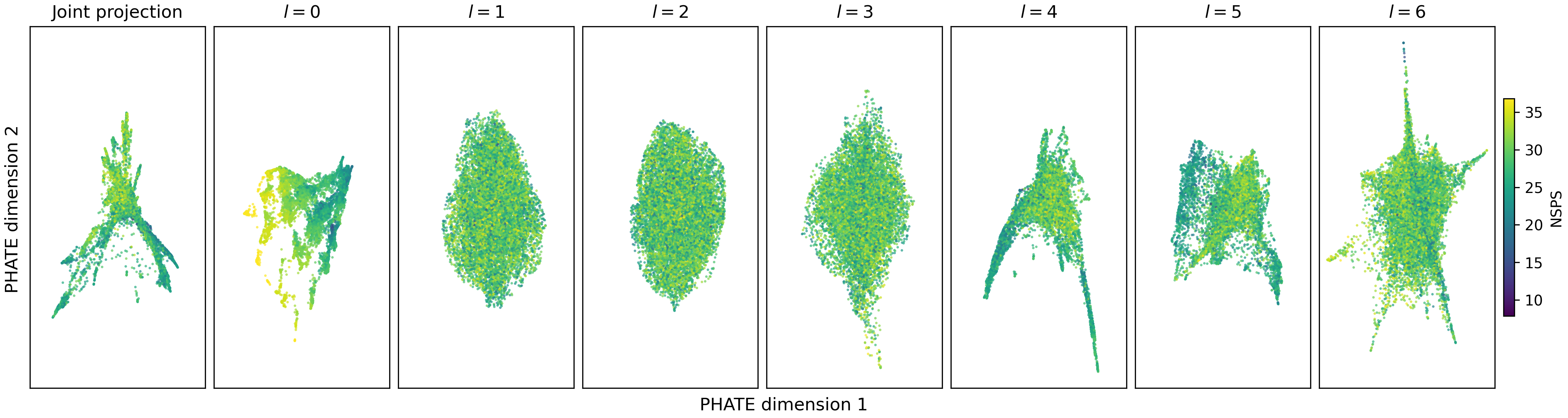}
        \caption{Embeddings produced for $L = [0,1,2,3,4,5,6]$.}
        \label{subfig:conventional}
    \end{subfigure}
    \begin{subfigure}[c]{0.9\textwidth}
        \centering
        \includegraphics[width=\linewidth]{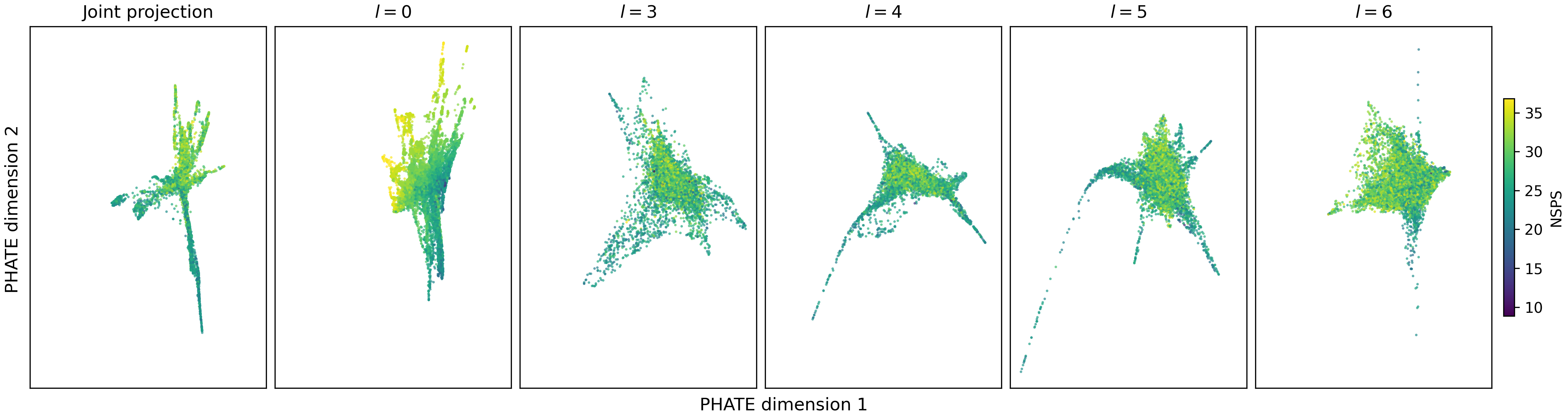}
        \caption{Embeddings produced for $L = [0,3,4,5,6]$. Note that the model was trained \emph{without} $l=1,2$.}
        \label{subfig:unconventional}
    \end{subfigure}
    \caption{\label{fig:main-plot} PHATE embedding projections for two configurations: for the same hidden dimension ($h=16$) and trained for the same number of epochs, \ref{subfig:conventional} uses a contiguous basis, while \ref{subfig:unconventional} skips $l=1,2$ in favor of adding $l=3,4,5,6$. The leftmost panel shows the PHATE projection when considering unified embeddings; from left to right, we decompose the unified embeddings into feature spaces that correspond to specific irreducible representations.}
\end{figure}

Figure \ref{fig:main-plot} compares PHATE projections of embeddings in the test set obtained with the two best performing models in the second half of Table \ref{tab:modelperformance}, namely $L=[0,1,2,3,4,5,6]$ and $L=[0,3,4,5,6]$. First and foremost, in both cases, the joint projections (i.e. embeddings from the final layer of dimensionality $h$) show some degree of structure, albeit noisy. The color distribution of individual points---each of which correspond to a molecular graph in the test split---shows that the learned joint representation does not clearly distinguish or ``recognize'' molecular complexity measured by NSPS, particularly in Figure \ref{subfig:conventional}. The joint projection in Figure \ref{subfig:unconventional}, appears to be slightly less noisy, and to a certain extent, there is a clearer partitioning of molecular complexity, with higher NSPS scores at the top than at the bottom.

It is in the projections of $l=0$ that seems to contain the most structure and information: the embeddings are significantly less noisy, i.e. less uniformly distributed, and present some relative ordering based on NSPS, to the extent that subclusters are visible particularly in $l=0$ of Figure \ref{subfig:conventional}. Within the same row, we see that $l=1,2,3$ embeddings amorphous, in stark contrast to $l=0$, but embedding structure re-appears at higher order $l$. In Figure \ref{subfig:unconventional}, the $l=3$ embeddings are noisy, but are significantly more structured than in Figure \ref{subfig:conventional}. Of particular interest are $l=4$ and $l=5$, where there is clear branching to extrema which can be interpreted as molecules that have maximal response to these two particular representations.\footnote{Refer to Section \ref{sec:branching} and Figure \ref{fig:mol-unconventional} for further discussion.}

With the combined context of Table \ref{tab:modelperformance} and Figure \ref{fig:main-plot}, we correlate the test generalization performance with the degree of semanticity or structure contained in the resulting embeddings. Figure \ref{subfig:conventional} shows comparatively less structure than Figure \ref{subfig:unconventional}, and results in the poorer model performance seen in Table \ref{tab:modelperformance}---we attribute this to how embeddings in $l=1,2,3$ are effectively noise, which dominates the scalar readout and hampers model efficacy.\footnote{This corroborates with how the joint projection embeddings are significantly less structured than $l=0$, which represents a very small fraction of the full embedding dimensionality. See Section \ref{sec:omitting-lower} for further discussion.} This behavior resembles that of early autoencoding generative models where the decoder is capable of minimizing the training loss without needing efficient and semantic embeddings---the encoder is scarcely updated during training and overwhelmingly resembles priors. This appears to be the $l=1,2,3$ results in Figure \ref{subfig:conventional}, and indeed, in other experiments (Figure \ref{fig:canonical-comparisons}). We conclude, then, that equivariant model training is still potentially capable of ignoring carefully crafted, physically inspired priors and latents, and persists even with longer training (\Cref{sec:longer-training}). Given the simple task of energy prediction, which is inherently a scalar and invariant quantity, the lack of embedding structure in $l=1,2,3$ in Figure \ref{subfig:conventional} shows that  equivariant models still fall into the common pitfall of maximum likelihood learning by overfitting certain features. On the one hand, basis functions by definition are convergent but if unused, are essentially wasted computation. We postulate that this may be remedied in a few ways:

\begin{itemize}
    \item Conventional regularization---dropout and other methods of encouraging weight sparsity may decrease the propensity of overfitting to specific irreducible representations.
    \item Tuning sets of spherical harmonics---Figure \ref{subfig:unconventional} interestingly shows that, by omitting intermediate orders $l=1,2$, the higher order terms actually \emph{gain} structure in their respective spaces. The contrast is particularly stark for $l=3$, which changes from being relatively amorphous to being less uniform how complexity in molecules are distributed in the latent space. When trained for the full 100 epochs, this configuration also becomes the best performing model (last row in Table \ref{tab:modelperformance}). Potentially, by pruning the feature space (the number of irreducible representations) and adapting the basis to data may result in improvements to modeling and computational performance.
    \item Pretraining on equivariant tasks---as previously mentioned, energy is an invariant property and therefore equivariance is not actually strictly required. While atomic forces require equivariance, when treated as a derivative of the energy it is unclear as to whether or not it constitutes as a strong \emph{training signal} in backpropagation that necessitates higher order tensor representations. Recent work has demonstrated the efficacy of denoising curricula for training both invariant and equivariant \citep{liaoGeneralizingDenoisingNonEquilibrium2024} models; pretraining from vector and/or tensor quantities directly may be necessary to guarantee utilization of features in each irreducible representations uniformly.
\end{itemize}

\section{Conclusions}

In this work, we adapt the improved spherical harmonics kernels developed in earlier work to understand how equivariant models use their spherical harmonic basis. We implement a simple equivariant graph convolution model and performed experiments on atomization energy prediction using the QM9 dataset, primarily treating the orders of spherical harmonics $L$ as a hyperparameter. A joint quantitative and qualitative perspective into the latent space yield two important, and perhaps largely counterintuitive insights: irreducible representations can be largely ignored and can degrade test performance, and the choice of $L$ can behave more like a hyperparameter---i.e. tuned---rather than a convergent basis.

It remains to be seen how these results transfer to other equivariant architectures and tasks---we have shown that it is \emph{possible} for physical latents to be ignored, in the same way as highly expressive decoders have been known to do in autoencoders. Our work demonstrates an analysis methodology that can help identify under/unused representations, which we hope will spur interest in subsequent analysis and design of equivariant models. As future work, we have proposed a number of potential remedies and experiments that may encourage improved efficiencies, both in terms of crafting pretraining curricula based on learning high rank properties, as well as the potential for pruning unused representations.

\bibliography{main}
\bibliographystyle{unsrtnat}

\appendix

\section{Appendix}

\subsection{Equivariance, spherical harmonics, \& angular momentum}

In this section, we give interested readers a brief background overview of the relationship between the spherical harmonic basis functions and angular momentum. The intention is to provide a bridge between deep learning practitioners (i.e. applications) where equivariance is a fairly recent idea, and quantum mechanics and spectroscopy, where these ideas have been developed and over the last century---many of which are described in seminal texts by \citet{zareAngularMomentumUnderstanding1988}, \citet{Rose2003-fr}, and with conventions set as far back as 1935 by \citet{condonTheoryAtomicSpectra1991}. In quantum mechanics, the general treatment of any system is often done with matrix representations: we describe a system in terms of some basis functions $\psi$ indexed by $i,j$, and a (Hamiltonian) operator $H$ determines some property of the system by solving $\langle \psi_i \vert H \vert \psi_j \rangle$\footnote{This notation is referred to as ``bra-ket'' notation, which effectively acts in this context as a short form for matrix multiplication with $\psi_{i,j}$ being row and column vectors. These vectors are also referred to as ``states'' in spectroscopy, which reflect vectors that provide a complete description of the system of interest; electron or nuclear spin, external electric or magnetic fields, etc.}---concretely, this means evaluating matrix elements and diagonalizing to obtain eigenvalues and eigenvectors, with the former corresponding to the actual quantities of interest. Note that this not only provide a means to numerically reach solutions, but bestows a number of desirable properties of $H$ and $\psi$ themselves when chosen correctly: commutation properties of $H$ allow derived quantities to be invariant or equivariant, while basis functions can be chosen deliberately to make matrices diagonal or at least block diagonal, providing a better abstraction and decreasing the need for computation. Thus, the relationship between spherical harmonics ($Y_{lm}$) and angular momentum operators ($\mathbf{L}$ for orbital angular momentum, and $\mathbf{J}$ for total angular momentum) is simply that $Y_{lm}$ are eigenfunctions of $\mathbf{L}^2$ and thus renders the matrix diagonal \citep{Rose2003-fr}.

The relevance with deep learning is to take advantage of these properties, as remarked by \citet{smidtFindingSymmetryBreaking2021}. Equivariance in physical systems is a consequence of angular momentum conservation rules, and the preservation of angular momentum necessitates the understanding of how to two or more sources of angular momentum \emph{couple} together---within the context of this equivariant neural networks, how two sets of feature vectors of particular irreducible representations with $d(2l +1)$ elements interact together and mix. The algebra of this is relatively straightforward \citep{Rose2003-fr}---in the so-called ``uncoupled'' representation, two sources of angular momentum denoted with $l_1,m_1$ and $l_2,m_2$ are described by basis functions $\psi_{l_1m_1}$ and $\psi_{l_2m_2}$ and operators $\mathbf{L_1}$ and $\mathbf{L_2}$; as previously described, they are diagonal in a basis of spherical harmonics. The ``coupled'' regime, with operator $\mathbf{L} = \mathbf{L_1} + \mathbf{L_2}$, has its representation $\psi_{lm}$ related to the uncoupled representation via the unitary transformation:

$$\psi_{lm} = \sum_{m_1m_2} C(l_1l_2l; m_1 m_2 m) \psi_{l_1m_1} \psi_{l_2m_2}$$

where $C(l_1l_2l; m_1 m_2 m)$ are the so-called Clebsch-Gordon coefficients, with the double summation occurring over $m_1 m_2$ projections. The Clebsch-Gordon coefficient can be seen to gate interactions---not all angular momentum coupling schemes between $l_1m_1$ and $l_2m_2$ are allowed, which gives rise to a number of simplifications and/or ``selection rules'' where $C$ is zero. Some are straightforward, such as the general conservation of angular momentum; $\vert l_1 - l_2 \vert \leq l \leq l_1 + l_2$ for $m = m_1 + m_2$. How $C$ is obtained is detailed in a number of texts, but concretely, $C$ is generally a orthogonal, real matrix of shape $(2l_1 + 1)(2l_2 + 1) \times (2l_1 + 1)(2l_2 + 1)$, many elements of which are zero as the condition $m = m_1 + m_2$ is not satisfied \citep{zareAngularMomentumUnderstanding1988}. As to the physical/geometric intuition of coefficients of $C^2$---the squared coefficients---they correspond to the probability density of interaction; for those more familiar with deep learning, this is akin to an attention mechanism as their values range from $[0,1]$. 

\begin{figure}
    \centering
    \includegraphics[width=\linewidth]{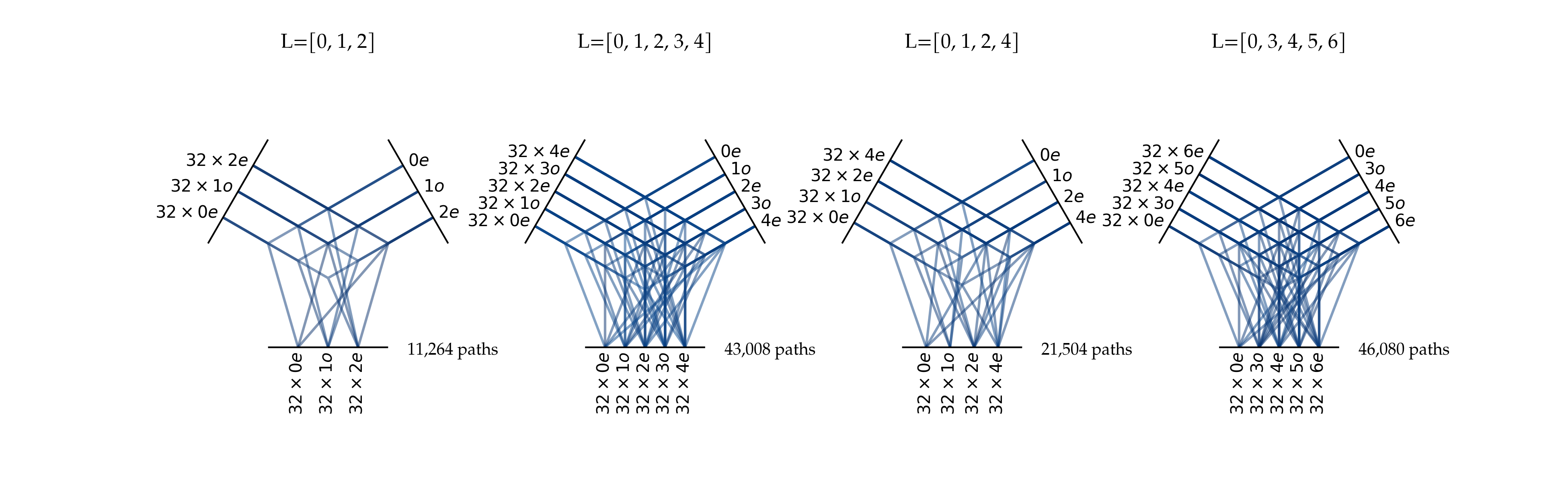}
    \caption{Tensor product paths for configurations of $L$ considered in this work. Input feature representations are shown in the top left of each diagram, with spherical harmonics on the right and outputs on the bottom. Here, the input features are assumed to be the output of the first interaction layer, i.e. we have already transformed the scalar atomic features. }
    \label{fig:tensor-paths}
\end{figure}

In \texttt{e3nn}, the ``valid'' tensor paths for tensor products correspond to non-zero probability amplitudes. The relevance to this work is such that there is no strict requirement that $L$ contains a contiguous set of spherical harmonic orders: providing that there are valid coupling schemes/paths that connect a pair of irreducible representations---e.g. conserve angular momentum---we can arbitrarily compose $l$ orders to form $L$. Figure \ref{fig:tensor-paths} uses \texttt{e3nn} to visualize allowed paths for a select few experimental configurations. In particular, $L=[0,1,2,4]$ skips $l=3$ but still includes paths that ensure that every irreducible representation is used, and in $L=[0,3,4,5,6]$, the number of paths are quite similar to $L=[0,1,2,3,4]$ despite the former having a higher order number.

\subsection{EquiTriton kernels}
\label{sec:equitriton}

\begin{figure}
    \centering
    \includegraphics[width=0.6\linewidth]{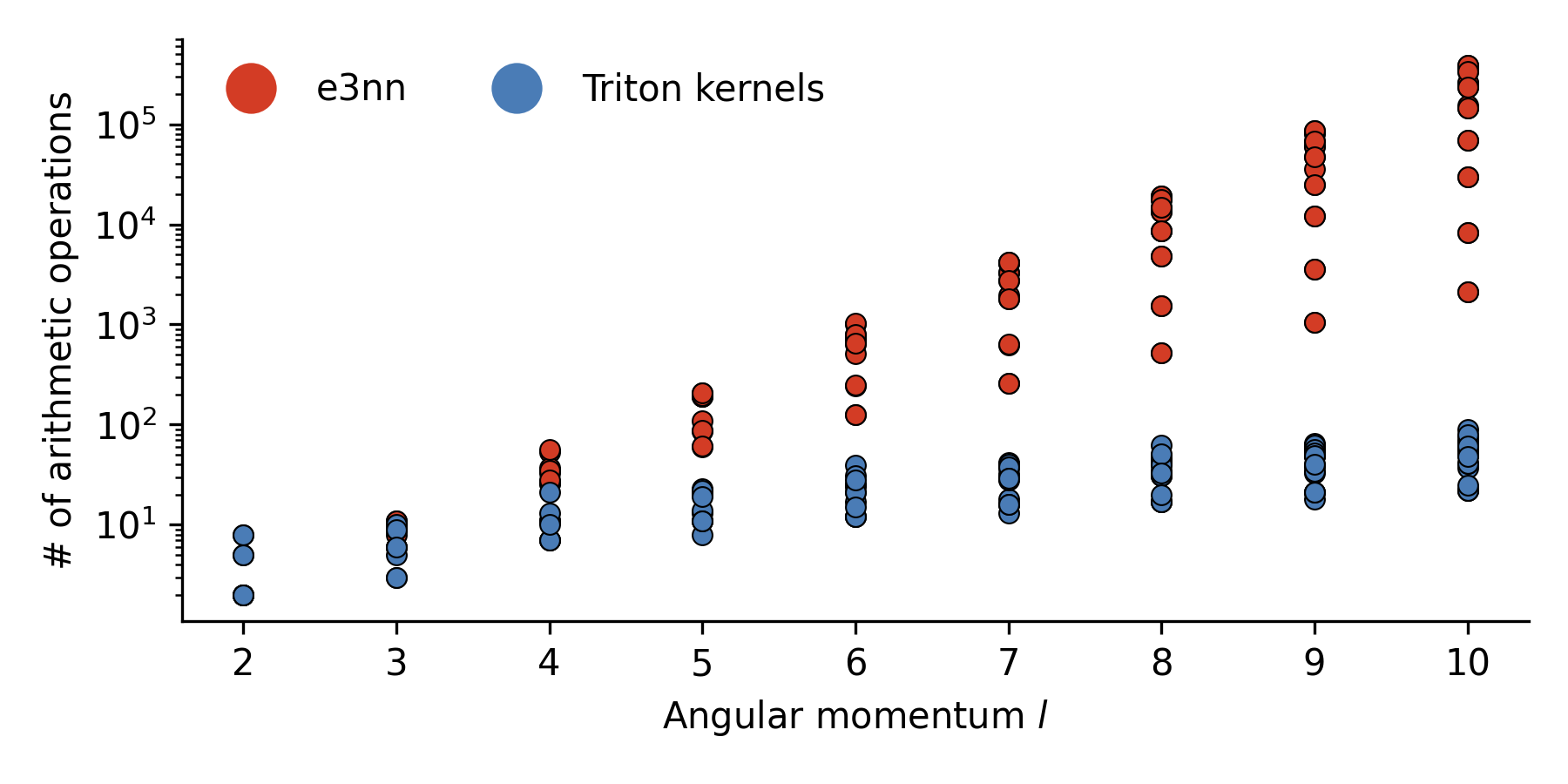}
    \caption{Visualization of the reduction in the number of arithmetic operations owing to aggressive symbolic refactoring. Each scatter point represents the number of arithmetic operations to compute a particular spherical harmonic $Y_{lm}$ for projection $m$ and order $l$. Red points correspond to a naive recurrent computation (i.e. higher $l$ depends on prior terms of $l$); blue points correspond to expressions derived and implemented in this work.}
    \label{fig:algoscaling}
\end{figure}

Building on top of earlier work \citep{leeScalingComputationalPerformance2024} in implementing performant kernels for spherical harmonics embedding, here we implemented a new set of kernels with significantly more aggressive symbolic refactoring. The main focus for the refactoring was to decouple each order of spherical harmonic: for every spherical harmonic order $l$ and its set of $2l+1$ projections, we used \texttt{sympy} \citep{sympy} to reformulate them as direct functions of $x,y,z$, removing the recurrent nature where terms in $Y_{lm}$ depend on $Y_{l-1 m}$. By removing this dependency, we are able to freely compose $L$ sets without impacting performance negatively. Furthermore, for every $Y_{lm}$, we performed a set of symbolic refactoring and ordering prior to reducing terms to constants, which significantly reduces the number of arithmetic operations as shown in Figure \ref{fig:algoscaling}. The number of operations are theoretically determined by \texttt{sympy}, and are not measured. We see that the number of operations is nearly constant with $l$ with the refactored expressions, while the cost of recurrence is clearly seen in the original implementation.

We have ensured that these kernels preserve equivariance using \texttt{e3nn} functionality, e.g. random rotation of coordinates, and checking that the embeddings produced are equivalent to direct rotation of the embeddings themselves. The remaining potential trade-off then is numerical accuracy and stability: while we have taken care to provide constant terms with sufficient digits of precision (up to double floating point precision), there may still be finite differences from the \texttt{e3nn} implementation due to the conversion of terms into rationals via \texttt{sympy}, rather than platform-specific compilers\footnote{At runtime, irrational values such as $\sqrt{37}$ are likely to be converted into literals---what this literal is exactly depends on the hardware platform and compiler, and depends on whether the operations are deterministic.}. Indeed, even implementing the same kernels in pure PyTorch was found to introduce numerical differences in a large fraction of trials in half precision, less so at single precision, and finally effectively absent at double precision. Thus while the Triton kernels are \emph{functionally} equivalent, it does not necessarily guarantee perfect exchangablility---i.e. a model trained with \texttt{e3nn} kernels using EquiTriton for inference may yield different results, and for training, may lead to different minima. This is particularly of import for molecular dynamics simulations which demand long timestep integrations, where numerical instability and error propagation compounds over time. We intend to investigate this aspect of the Triton kernels more thoroughly in the near future.

\subsection{Model architecture \& training}

\label{sec:architecture}

We developed a heavily simplified equivariant graph neural network architecture for experiments detailed in this paper. The architecure is implemented using \texttt{e3nn}, with the option to interchange spherical harmonics kernels between Triton and \text{e3nn}'s own \texttt{torchscript} implementations. The overall model takes an input molecular graph, which comprises atomic numbers and positions for each node, and edges that describe connectivity between nodes. The architecture comprises a sequence of replicated interaction or equivariant graph convolution blocks. In the initial input layer, the atomic numbers are used to index an embedding table, mapping each atom type as scalar ($l=0$) features. The atom positions are embedded in two ways: pairwise \emph{distances} (scalar) are expanded in a basis of Bessel radial functions and passed into a multilayer perceptron, while pairwise displacements (vectors) are mapped onto the set of spherical harmonic functions. A fully connected tensor product combines---whilst preserving their respective representations---the scalar atomic features, radial features, and spherical harmonics to yield messages, and a scatter add subsequently updates node features.

The model was implemented in PyTorch \citep{paszkePyTorchImperativeStyle2019a} using PyTorch Geometric \citep{feyFastGraphRepresentation2019a} for message passing abstraction, and PyTorch Lightning \citep{Falcon_PyTorch_Lightning_2019} for abstracting accelerator offloading, training, and evaluation. The \texttt{e3nn} \citep{e3nn} was used for computing tensor products. For all of our experiments, we kept hyperparameters fixed except for those mentioned in the main text---\Cref{tab:hyperparameters} shows the general configuration used.

\begin{table}[]
    \centering
    \caption{General hyperparameters used for model training. The training seed is passed into the PyTorch Lightning \texttt{seed\_everything}, which sets the same seed for PyTorch, NumPy, and builtin \texttt{math} libraries. The degree normalization term is used to rescale message updates.}
    \begin{tabular}{c|c}
        \toprule
        Parameter & Value \\
        \midrule
         Optimizer & \texttt{AdamW} \\
         Learning rate & $10^{-3}$ \\
         Random seed & $21616$ \\
         Train/val./test split fraction & $0.8/0.1/0.1$ \\
         Label normalization ($\mu/\sigma$) & $-76.116/10.3238$ \\
         Degree normalization & $6.0828$ \\
         Initial atom embedding dim. & $64$ \\
         \# of bessel functions & $20$ \\
         Radius cutoff & $6.0$ \\
         \# of interaction blocks & $3$ \\
         \bottomrule
    \end{tabular}
    \label{tab:hyperparameters}
\end{table}

\subsection{Spacial scores for molecular complexity}
\label{sec:nsps}

The PHATE projections are colored throughout this paper using the normalized spacial score (NSPS) \citep{krzyzanowskiSpacialScoreComprehensive2023}, which provides a easily calculable metric for quantifying the relative complexity of a molecule. The metric is defined as $\mathrm{NSPS} = \frac{\sum^N_{n=1} sp_n s_n r_n H_n^2}{N_H}$ where the summation is over $N$ atoms in a molecule, with $sp$ the degree of hybridization (e.g. three for $sp^3$), $s$ is the number of stereoisomers (2 for possible $E/Z$, 1 otherwise), $r$ being a measure of aromaticity or linearity ($r=1$ if so, otherwise 2), and finally, $H$ as the number of heavy atom neighbors. The normalization factor $N_H$ corresponds to the total number of heavy atoms in the molecule, which accounts for the size of a molecule. Intuitively, large NSPS scores indicate complex motifs in a molecule; relevant to QM9, this generally pertains species that are highly branched and saturated (i.e $sp^3$ hybridized atoms), and to a lesser extent aromatic rings.

\subsection{Additional experiments and ablations}

\subsubsection{Using higher order spherical harmonics while omitting lower orders}

\label{sec:omitting-lower}

\begin{figure}
    \centering
    \begin{subfigure}[c]{0.9\textwidth}
        \includegraphics[width=\linewidth]{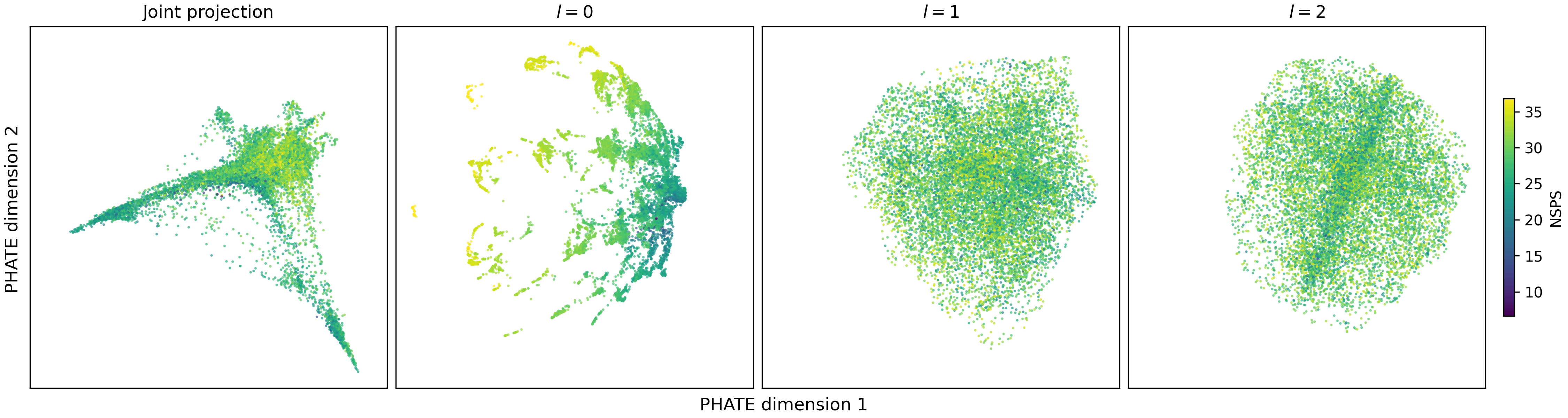}
        \caption{Embeddings from $L=[0,1,2]$, with a hidden dimension of 32.}
    \end{subfigure}
    \begin{subfigure}[c]{0.9\textwidth}
        \includegraphics[width=\linewidth]{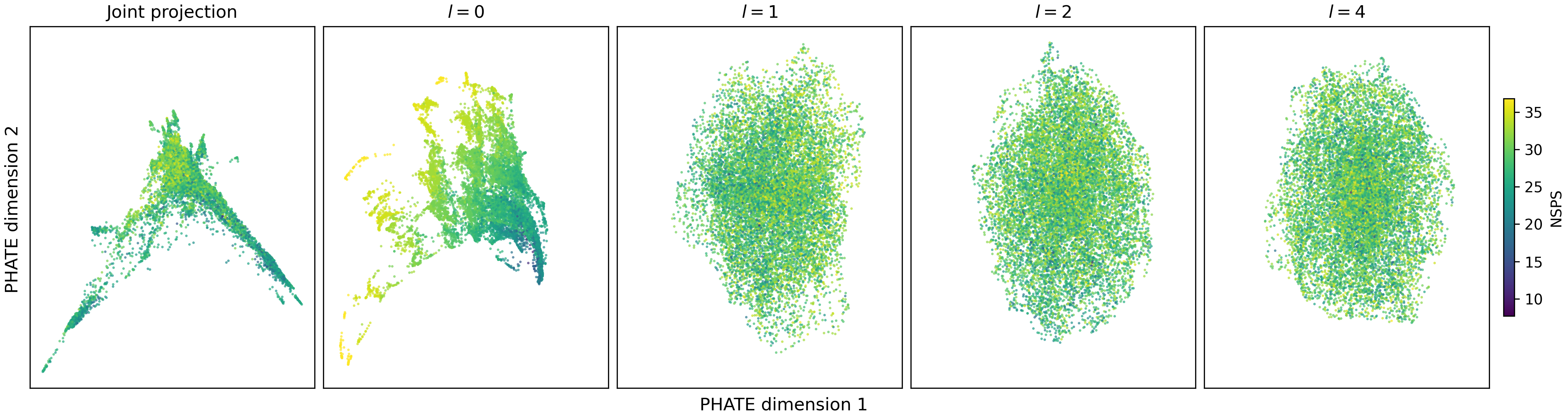}
        \caption{Embeddings from $L=[0,1,2,4]$, with a hidden dimension of 32.}
    \end{subfigure}
    \begin{subfigure}[c]{0.9\textwidth}
        \includegraphics[width=\linewidth]{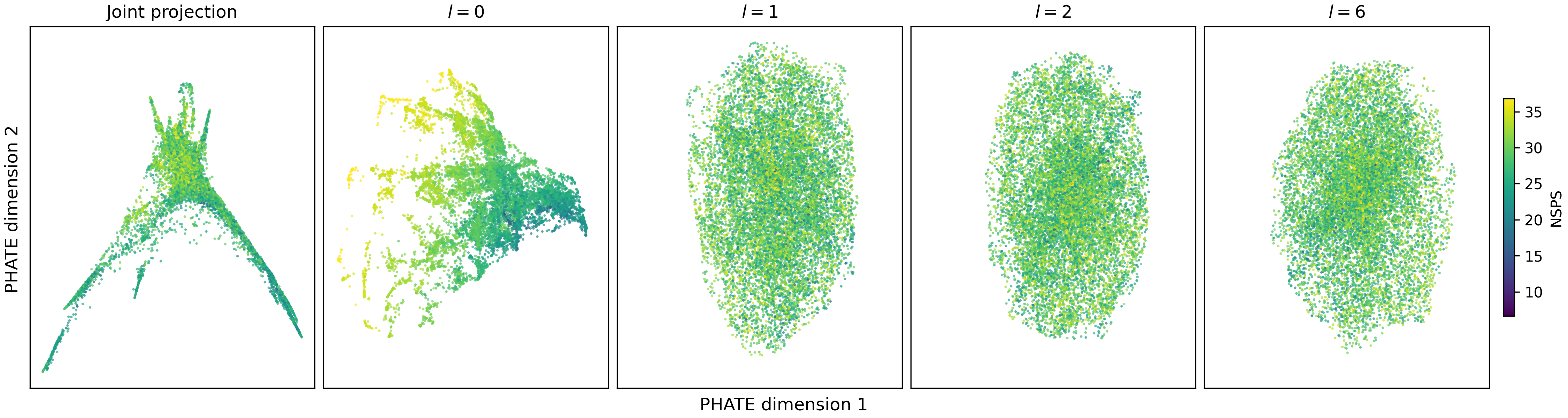}
        \caption{Embeddings from $L=[0,1,2,6]$, with a hidden dimension of 32.}
    \end{subfigure}
    \begin{subfigure}[c]{0.9\textwidth}
        \includegraphics[width=\linewidth]{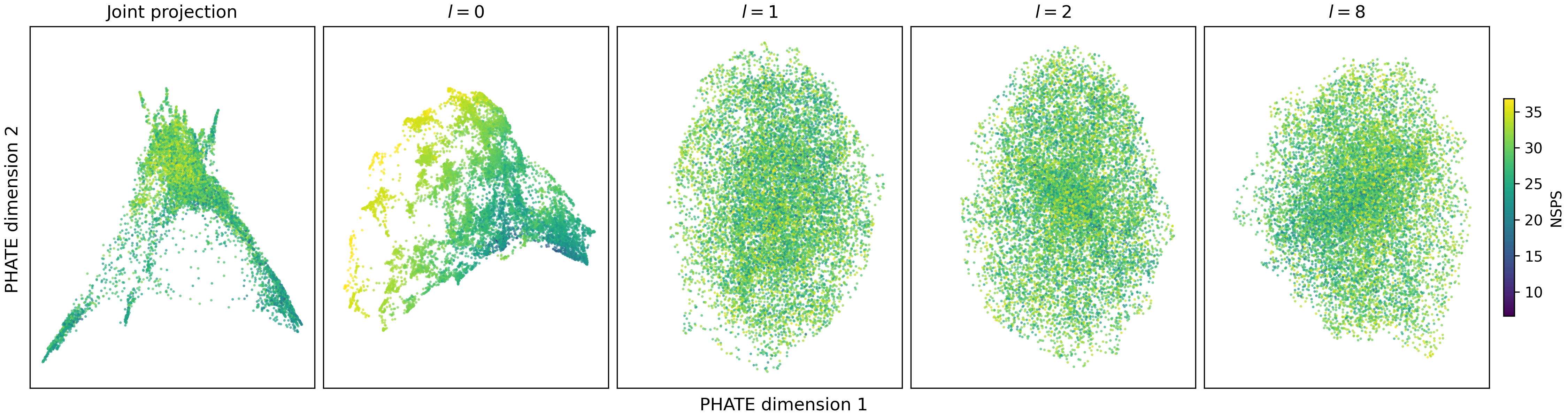}
        \caption{Embeddings from $L=[0,1,2,8]$, with a hidden dimension of 32.}
    \end{subfigure}
    \begin{subfigure}[c]{0.9\textwidth}
        \includegraphics[width=\linewidth]{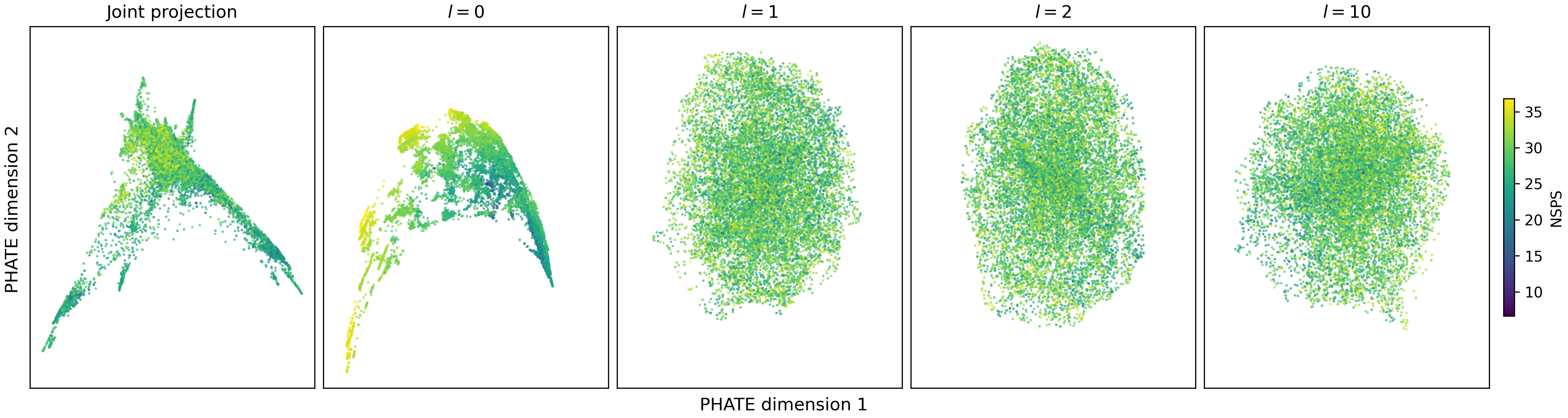}
        \caption{Embeddings from $L=[0,1,2,10]$, with a hidden dimension of 32.}
    \end{subfigure}
    \caption{PHATE embedding projections when considering the ``canonical'' equviariant set of spherical harmonics ($l=0,1,2$). With the exception of the first row, we include an one additional set of higher order spherical harmonics of even parity, increasing in orbital angular momentum from top to bottom.}
    \label{fig:canonical-comparisons}
\end{figure}

Figure \ref{fig:canonical-comparisons} shows PHATE projections for experiments in the top half of Table \ref{tab:modelperformance}, where we observed degradation of test performance of models when we append a higher order spherical harmonic of even parity to the canonical set of $L=[0,1,2]$. The embedding projections provide a rationale for this, as $l \neq 0$ does not present any embedding structure for any of the configurations, we are essentially adding $h(2l+1)$ elements of noise to the regression model, and $l=0$ is the only representation that consistently shows structure. From this perspective, the canonical set understandbly provides the best performance, as the effective signal-to-noise ratio (i.e. $l=0$ versus $l\neq0$) is the highest.

\subsubsection{Considering a contiguous $L$ set up to $l=4$}

\begin{figure}[h]
    \centering
    \includegraphics[width=\linewidth]{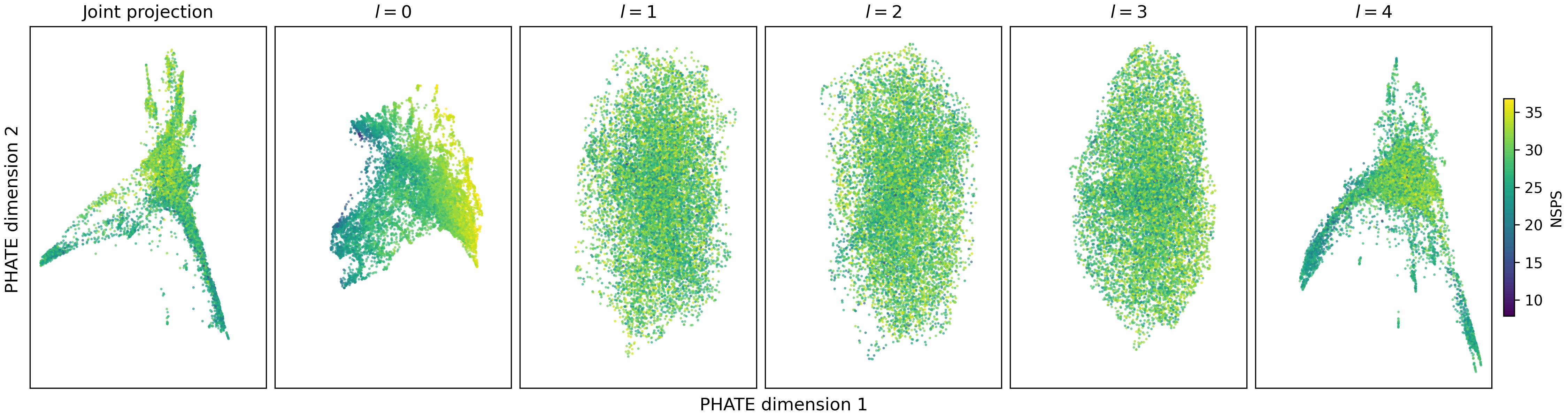}
    \caption{PHATE projections for a contiguous set of $L=0,1,2,3,4$, with a hidden dimension of 16. This is directly comparable and consistent with Figure \ref{subfig:conventional}: $l=1,2,3$ do not contain structure in their embeddings, but $l=4$ appears to contain information.}
    \label{fig:nano}
\end{figure}

From Figure \ref{subfig:conventional}, it is not immediately clear whether the structure arising from $l\geq4$ occurs by chance---i.e. whether the specific set of hyperparameters leads to $l=1,2,3$ being unstructured, and $l\geq4$ only happens to end up gaining structure. Figure \ref{fig:nano} shows PHATE projections for a similar experiment where we consider a contiguous set of $L$, albeit truncating up to $l=4$. We observe the same lack of overall structure in $l=[1,2,3]$, which is consistent with Figure \ref{subfig:conventional}, and generally Figure \ref{fig:canonical-comparisons}. The latent structure appearing consistently in $l=4$ across our experiments does suggest that there is some degree of affinity with the molecular graphs, akin to a ``matched filter'' in signal processing. With this interpretation, $l=1,2,3$ do not produce a sizable signal when convolved (or rather, cross-correlated) with the data.

As to why this particular set, $L=[0,1,2,3,4]$ shows structure but not $L=[0,1,2,4]$ in Figure \ref{fig:canonical-comparisons}, as our current set of experiments do not provide evidence, we can only speculate that there may be that \emph{in this particular case}, $l=4$ depends on coupling with $l=3$, and/or the training dynamics resulted in converging on a point in the loss landscape that did not require the use of $l=4$. Additional experiments with techniques such as input gradients \citep{simonyanDeepConvolutionalNetworks2014} may provide the required insight into the relationship between $l$ orders.

\subsubsection{Embedding visualization with longer training}
\label{sec:longer-training}

\begin{figure}
    \centering
    \begin{subfigure}[c]{0.9\textwidth}
        \includegraphics[width=\linewidth]{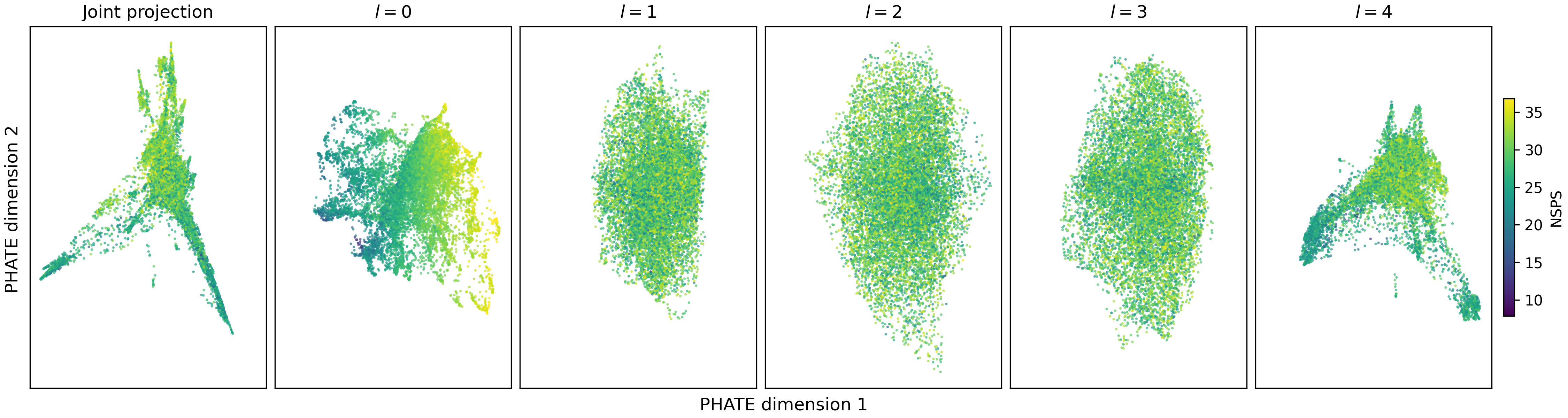}
        \caption{Embeddings from $l=[0,1,2,3,4]$, with a hidden dimension of 16 and trained for 100 epochs. These projections can be compared with \Cref{fig:nano}, which were trained for 30 epochs.}
    \end{subfigure}
    \begin{subfigure}[c]{0.9\textwidth}
        \includegraphics[width=\linewidth]{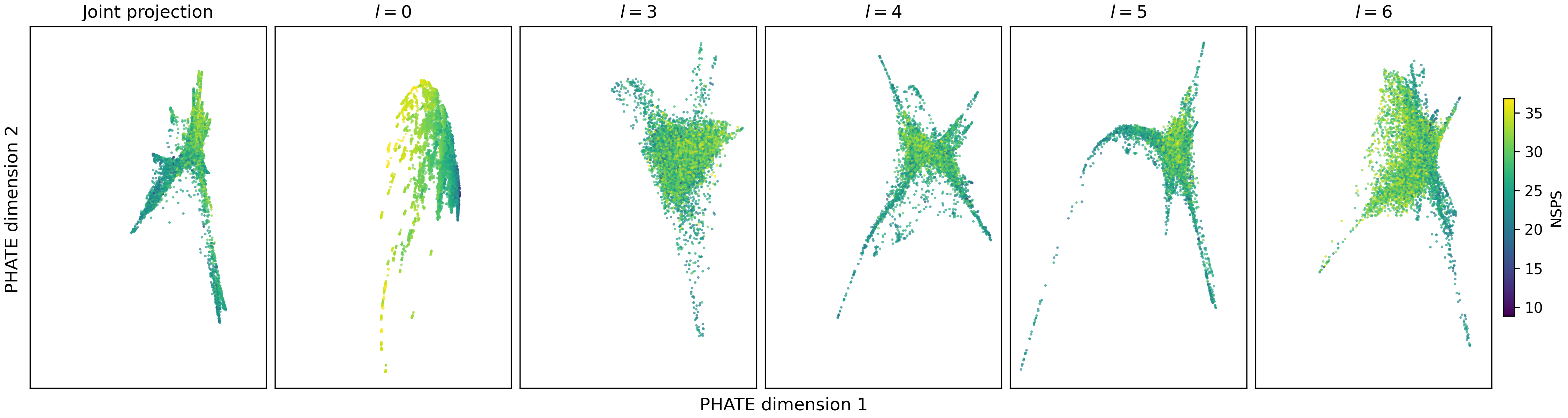}
        \caption{Embeddings from $l=[0,3,4,5,6]$, with a hidden dimension of 16 and trained for 100 epochs. These projections can be compared with \Cref{subfig:unconventional}, which was trained for 30 epochs.}
        \label{subfig:unconventional-long}
    \end{subfigure}
    \caption{PHATE embedding projections for two configurations in Table \ref{tab:modelperformance} with training up to 100 epochs.}
    \label{fig:longer-epochs}
\end{figure}

Figure \ref{fig:longer-epochs} shows the change in latent space with longer training: from 30 to 100 total epochs. In both experiments with $L$, the PHATE projections show that the latent spaces do not qualitatively change with more training. While the model performance improves for both in terms of the test metric, the fact that the latent spaces do not change significantly indicates that the \emph{global} structure is learned quite early on, with some small adjustments to local structure only. This result leads us to speculate that the latent structure could be largely dictated by strong physical priors that may not update well with data---at least when trained with backpropagation. From this perspective, equivariant models may not benefit \emph{as well as they could} from naive data and model scaling; the study by \citet{freyNeuralScalingDeep2023} does show data scaling for PaiNN \citep{schuttEquivariantMessagePassing2021}, but an embedding analysis is likely needed to ascertain whether $l\geq1$ embeddings are being used. Our result with $L=[0,3,4,5,6]$ shows room for improvement over the canonical $L=[0,1,2]$ baseline when the embedding space is being used.

\subsubsection{Branching and clusters in the latent space}
\label{sec:branching}

\begin{figure}[h]
    \centering
    \begin{subfigure}[c]{0.45\textwidth}
        \includegraphics[width=\linewidth]{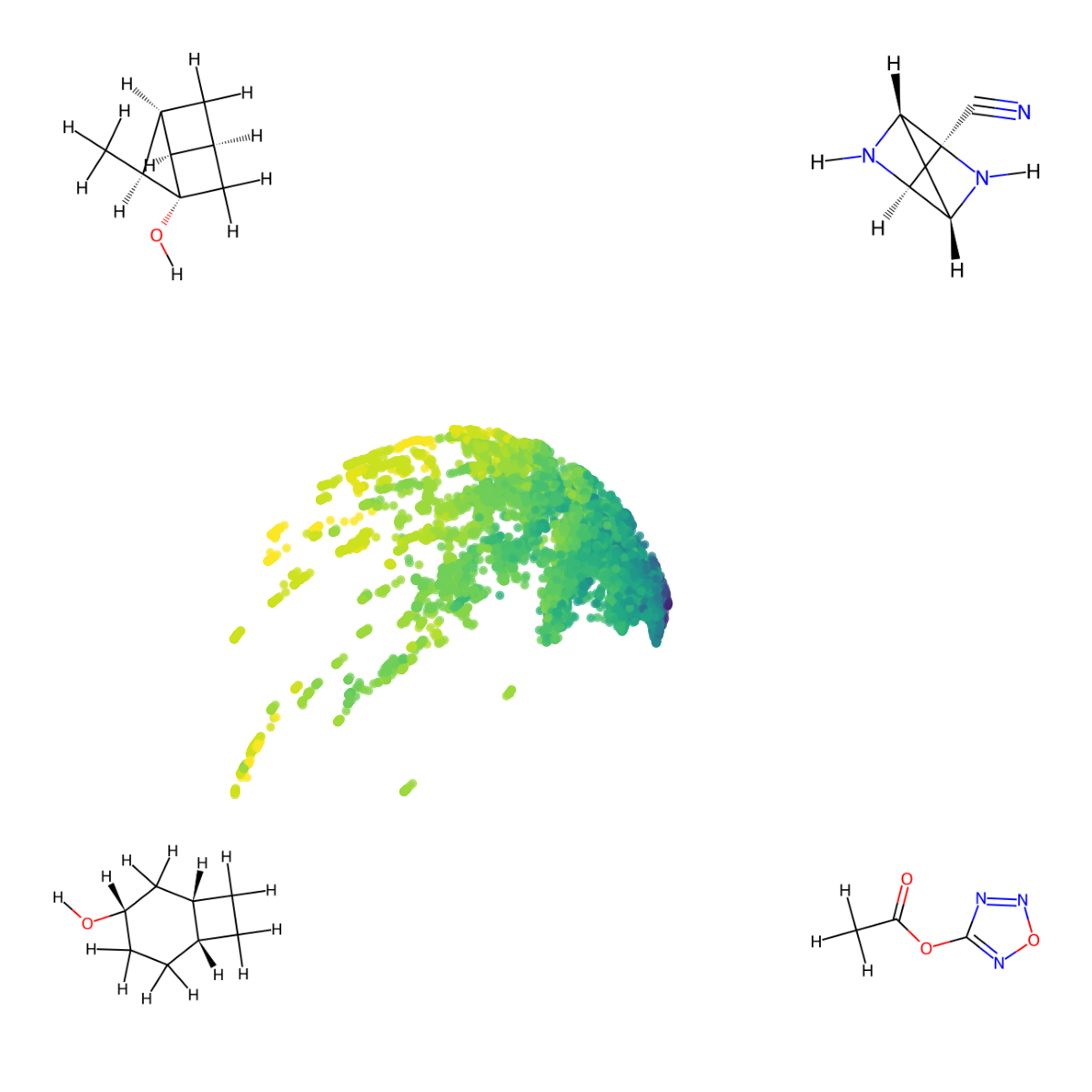}
        \caption{$l=0$}
    \end{subfigure}
    \begin{subfigure}[c]{0.45\textwidth}
        \includegraphics[width=\linewidth]{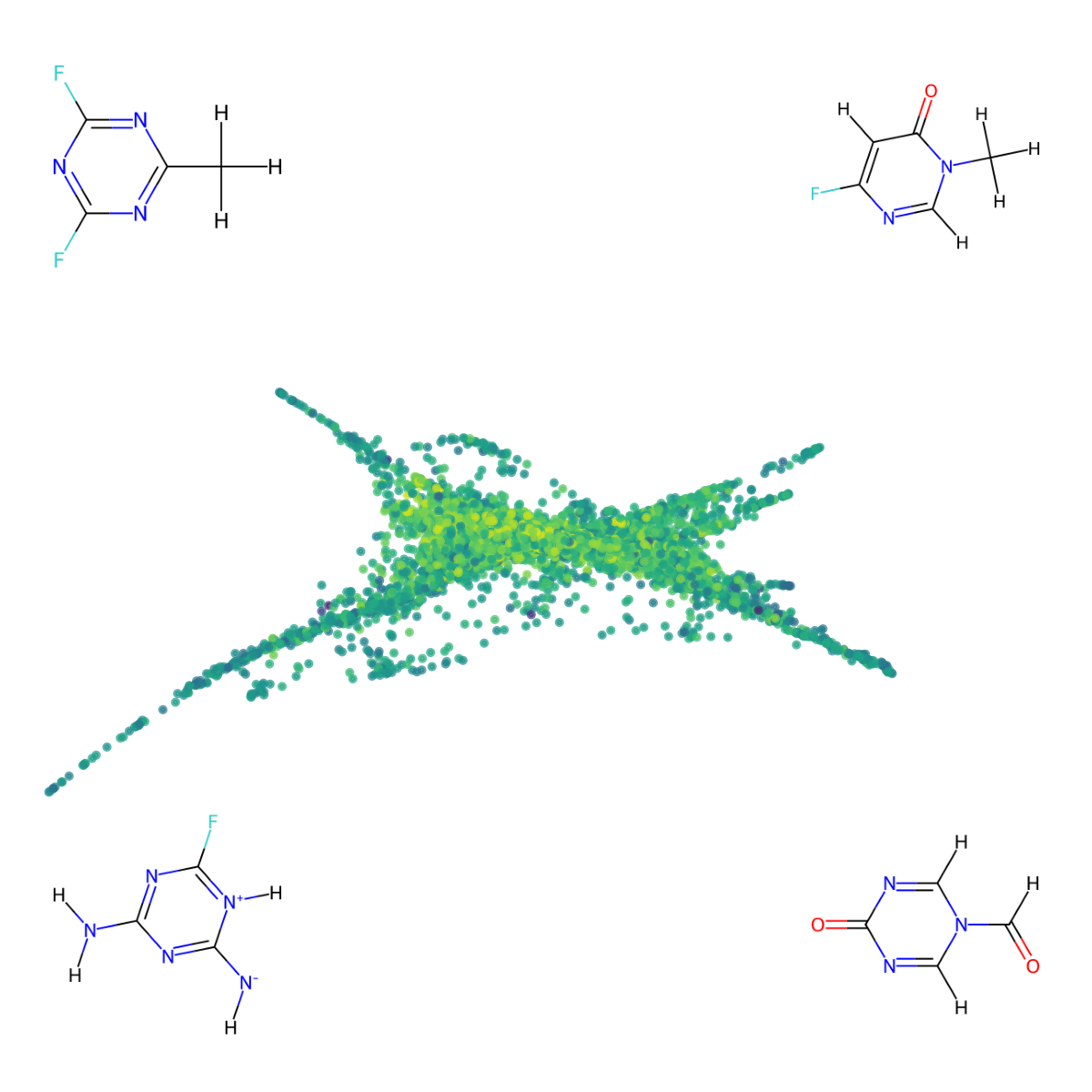}
        \caption{$l=4$}
    \end{subfigure}
    \caption{PHATE projections with molecule image annotations overlaid. Structures correspond to the nearest molecule in the embedding space to the edges (i.e. north-east, south-west) of the four quadrants. The projections are equivalent to Figure \ref{subfig:unconventional-long}, albeit with a different aspect ratio for visualization purposes.}
    \label{fig:mol-unconventional}
\end{figure}

In Section \ref{sec:embedding-projection} we alluded to the interpretation of branches in the PHATE projected global structure. As PHATE is capable of preserving both local and global manifold structure \citep{moonVisualizingStructureTransitions2019}, we can interpret aspects like branching and clustering with more flexibility than what would be afforded by methods like $t$-SNE and UMAP. With this in mind, and interpreting the action of the spherical harmonics as signal filters, Figure \ref{fig:mol-unconventional} shows the PHATE projections from Figure \ref{subfig:unconventional-long} with the most distance molecules found in each quadrant, which represent the extremes. This visualization is ideally suited for $l=4$, where there are conveniently four branches that extend into these quadrants, and allow us to speculate on what kind of molecules match well to $l=4$ weights. The general trend appears to differentiate between aromatic rings, and identifies four branches based on composition---roughly speaking, oxygen-bearing species on the right (east), and the degree of fluourination as well. In the scalar representation ($l=0$), the more saturated (i.e. $sp^3$ hybridized) structures branch towards the left (west) both from the example structures and from the NSPS color map.

\end{document}

%% file: math_commands.tex
\usepackage{amsmath,amsfonts,bm}
\usepackage{mathtools}

\def\eqref#1{equation~\ref{#1}}

\def\1{\bm{1}}

\DeclareMathAlphabet{\mathsfit}{\encodingdefault}{\sfdefault}{m}{sl}
\SetMathAlphabet{\mathsfit}{bold}{\encodingdefault}{\sfdefault}{bx}{n}

\def\1{{\mathbf{1}}}